\documentstyle[12pt]{article}
\topmargin - 0.8cm

\catcode`@=11
\@addtoreset{equation}{section}
\catcode`@=12

\begin{document}
\title{\vskip-1.7cm \bf  Quantum scale of inflation and particle
physics of the
early universe}
\author{A.O.Barvinsky$^{1,2}$\ and A.Yu.Kamenshchik$^{2}$ }
\date{}
\maketitle
\hspace{-8mm}$^{1}${\em
Theoretical Physics Institute, Department of Physics, \ University of
Alberta, Edmonton, Canada T6G 2J1}
$^{*}$
\\ $^{2}${\em Nuclear Safety
Institute, Russian Academy of Sciences , Bolshaya Tulskaya 52, Moscow
113191, Russia}
\begin{abstract}
The quantum gravitational scale of inflation is calculated by finding
a sharp
probability peak in the distribution function of chaotic inflationary
cosmologies driven by a scalar field with large negative constant
$\xi$ of
nonminimal interaction. In the case of the no-boundary state of the
universe
this peak corresponds to the eternal inflation, while for the
tunnelling
quantum state it generates a standard inflationary scenario. The
sub-Planckian
parameters of this peak (the mean value of the corresponding Hubble
constant
${\mbox{\boldmath $H$}}\simeq 10^{-5}m_P$, its quantum width
$\Delta{\mbox{\boldmath $H$}}/{\mbox{\boldmath $H$}}\simeq 10^{-5}$
and the number of inflationary
e-foldings ${\mbox{\boldmath $N$}}\simeq 60$) are found
to be in good correspondence with the observational status of
inflation theory,
provided the coupling constants of the theory are constrained by a
condition
which is likely to be enforced by the (quasi) supersymmetric nature
of the
sub-Planckian particle physics model.
\end{abstract}
\\
$^{*}$ Present address\\

\baselineskip6.8mm
\section{Introduction}
\hspace{\parindent}
It is widely reckognized that one of the most promising pictures of
the early
universe is a chaotic inflationary scenario \cite{Linde}. The
inflation
paradigm is the more so attractive that it allows to avoid the
fortune telling
of quantum gravity and cosmology because the inflationary epoch is
supposed to
take place at the energy scale or a characteristic value of the
Hubble constant
$H=\dot a/a \sim 10^{-5}m_P$ much below the Planck one $m_P=G^{1/2}$.
However,
the predictions of  the inflation theory essentially depend on this
energy
scale which must be chosen to provide a sufficient number of
e-foldings $N$ in the expansion law of a scale factor $a(t)$ during
the
inflationary epoch, $N=\int _{t_{\rm in}}^{t_{\rm fin}} dt\,H={\rm
ln}\,(a_{\rm
fin}/a_{\rm in})$,
and also generate the necessary level of density
perturbations capable of the formation of the large scale structure.
In the
chaotic inflationary model the Hubble constant  $H=H(\phi)=\sqrt{8\pi
U(\phi)/3m_P^2}$ is effectively generated by the potential $U(\phi)$
of the
inflaton scalar field $\phi$ which satisfies the slow-roll
approximation
\cite{Linde},
	$\dot\phi\simeq -(1/3H)\partial U/\partial\phi\ll H\phi$.
The number of e-foldings $N=N(\phi_I)$, the effective Hubble constant
$H=H(\phi_I)$ and the generation of the density perturbations, as
well as the
validity of the slow-roll approximation itself,  essentially depend
upon one
parameter -- the initial value of the inflaton field $\phi_I$, and
one of the
most fundamental observational bounds is the following restriction on
this
quantity \cite{Linde}
	$N(\phi_I)\simeq (4\pi/m_P^2)\int_{0}^{\phi_I}d\phi\,H(\phi)
	\Big[\partial H(\phi)/\partial\phi\Big]^{-1} \geq 60 $.

This quantity, however, is a free parameter in the inflation theory,
and, to
the best of our knowledge, there are no convincing principles that
could fix it
 without invoking the ideas of quantum gravity and cosmology.  These
ideas
imply that there exists a quantum state of coupled gravitational and
matter
fields, which in the semiclassical regime generates the family of
inflationary
universes with different values of $H(\phi)$, approximately evolving
at later
times according to classical equations of motion. This quantum state
allows one
to calculate the distribution function $\rho(\phi)$ of these
universes and
interprete its maximum at certain value of $\phi=\phi_I$ (if any) as
generating
the quantum scale of inflation. The implementation of this approach,
undertaken
in the tree-level approximation for the no-boundary \cite{HH,H} and
tunnelling
\cite{Vilenkin} quantum states of  the Universe, was not successful.
The
corresponding distribution functions turned out to be extremely flat
\cite{H-Page,Vilenkin:tun-HH} for large values of $\phi$ (in the
domain of the
inflationary slow-roll ansatz) and unnormalizable at
$\phi\rightarrow \infty$, which totally breaks the
validity of the semiclassical expansion underlying the inflation
theory, since
the contribution of the
over-Planckian energy scales is not suppressed to zero.  Apart from
this
difficulty,  the only local maximum of $\rho(\phi)$ found for the
no-boundary
quantum state was shown to be generating insufficient amount of
inflation
violating the above bound \cite{GrisR}.

In the series of recent papers \cite{BKam:norm,BarvU,tunnel} we
proposed the
mechanism of suppressing the over-Planckian energy scales by the
contribution
of the quantum (loop) part of the gravitational effective action.
This can
justify the use of the semiclassical expansion and serve as a
selection
criterion of physically viable particle models with the normalizable
quantum
state, suggesting the supersymmetric extension of field models in the
theory of the early universe \cite{Kam:super}. Here we further apply
this
mechanism to show that it can also generate the quantum scale of
inflation and,
in particular, serve as a quantum gravitational ground for the
inflationary
model of Bardeen, Bond and Salopek \cite{SalopBB} with large negative
constant
$\xi$ of nonminimal inflaton-graviton coupling.

\section{Probability of inflation with nonminimal inflaton field:
tree-level
approximation}
\hspace{\parindent}
Two known proposals for the cosmological quantum state, which
semiclassically
generate the families of inflationary universes,  are represented by
the
no-boundary \cite{HH,H} and tunnelling \cite{Vilenkin} wavefunctions.
They
describe two different types of nucleation of the Lorentzian
quasi-DeSitter
spacetime from its Euclidean counterpart  which in the context of
spatially
closed cosmology can be represented by the 4-dimensional Euclidean
hemishere
matched across the equatorial section to the Lorentzian expanding
Universe. The
tree-level no-boundary $\rho_{N\!B}(\phi)$ \cite{HH} and
tunnelling $\rho_{T}(\phi)$ \cite{rhoT} distribution functions of
such
universes are just the squares of their wavefunctions
	\begin{eqnarray}
	\rho_{N\!B}(\phi)\sim {\rm e}^{\textstyle -I(\phi)},\,\,\,
	\rho_{T}(\phi)\sim {\rm e}^{\textstyle +I(\phi)} ,
\label{2.1}
	\end{eqnarray}
where $I(\phi)$ is a doubled Euclidean action of the theory
calculated on such
a hemisphere (or the action on the full quasi-spherical manifold).
When $\phi$
belongs to the domain of the slow-roll approximation and is
practically
constant in the solution of both Lorentzian and Euclidean equations
of motion,
the Euclidean spacetime only slightly differs from the exact
4-dimensional
sphere of radius  $1/H(\phi)=\sqrt{3m_p^2/8\pi U(\phi)}$ -- the
inverse of the
Hubble constant -- and  $I(\phi)$ takes the form
	$I(\phi)=-3m_P^4/8U(\phi)$.
Thus,  $\rho_{NB}(\phi)$ and  $\rho_{T}(\phi)$ describe opposite
outcomes of
the most probable underbarrier penetration: respectively to the
minimum and to
the maximum of the inflaton potential $U(\phi)\geq 0$ (although, in
the former
case the minimum $U(\phi)=0$ falls out of the slow-roll domain ).

The equations above apply to the case of an inflaton field minimally
coupled to
the metric tensor $G_{\mu\nu}$ with the Lagrangian
$L(G_{\mu\nu},\phi)=G^{1/2}[\,(m_P^2/16\pi)\,R\,(G_{\mu\nu})-
(\nabla\phi)^2/2-U(\phi)\,]$, but can also be used in the theory of
the
nonminimal scalar field $\varphi$
	\begin{eqnarray}
	{\mbox{\boldmath $L$}}(\,g_{\mu\nu},\varphi\,)
	=g^{1/2}\left\{\frac{m_P^2}{16\pi}R\,(g_{\mu\nu})-
	\frac12\, \xi\varphi^2R\,(g_{\mu\nu})
	-\frac 12 \,(\nabla\varphi)^2
	-\frac12\,m^2\varphi^2-\frac{\lambda}{4}\,\varphi^4\right\} ,
\label{2.3}
	\end{eqnarray}
provided  $L(G_{\mu\nu},\phi)$ above is viewed as the Einstein frame
of
${\mbox{\boldmath $L$}}(\,g_{\mu\nu},\varphi\,)$ with the fields
$(G_{\mu\nu},\phi)=((1+8\pi|\xi|\varphi^2/m_P^2)g_{\mu\nu},
\phi(\varphi))$
related to $(\,g_{\mu\nu},\varphi\,)$ \cite{SalopBB,Page:conf,BKK}.
For a
negative nonminimal coupling constant $\xi=-|\xi|$  this model easily
generates
the chaotic inflationary scenario \cite{Spok-Unr} with the Einstein
frame
potential
	\begin{eqnarray}
	U(\phi)\,\Big|_{\,\textstyle\phi\!=\!\phi(\varphi)}=
	\frac{m^2\varphi^2/2+\lambda\varphi^4/4}
	{\Big(1+8\pi |\xi\,|\varphi^2/m_P^2\Big)^2} ,
      \label{2.4}
	\end{eqnarray}
including the case of a symmetry breaking at scale $v$ with
$m^2=-\lambda v^2<0$ in the Higgs potential
$\lambda(\varphi^2-v^2)^2/4$ . At large $\varphi$ it approaches
a constant and depending on the parameter
$\delta\equiv -8\pi|\xi|m^2/\lambda m_P^2=8\pi|\xi|v^2/m_P^2$ has two
types of
behaviour at the intermidiate values of the inflaton field. For
$\delta>-1$ it
does not have local maxima and generates the slow-roll decrease of
the
scalar field leading to a standard scenario with a finite
inflationary stage,
while for $\delta<-1$ it has a local maximum at
$\bar\varphi=m/\sqrt{\lambda|1+\delta|}$ and due to a negative slope
of the
potential leads to the eternal inflation for all models with the
scalar field
growing from its initial value  $\varphi_I>\bar\varphi$.

The tree-level distribution functions (\ref{2.1}) for such a
potential do not
suppress the over-Planckian scales and are unnormalizable at large
$\varphi$,
$\int^{\infty}d\varphi \,\rho_{N\!B,\,T}(\varphi) =\infty$, thus
invalidating
a semiclassical expansion. Only in the tunnelling case
with $\delta<-1$ the distribution $\rho_{T}(\phi)$ has a local peak
at $\bar\varphi$, which is physically unacceptable because for
reasonable
values $\xi=-2\times10^4$, $\lambda=0.05$ \cite{SalopBB} it generates
insufficient number of e-foldings $N\sim10^{-2}$ (for universes
sliding to the
left of the potential maximum at $\bar\varphi$) and also falls out of
the
domain of the slow-roll approximation.

\section{Beyond the tree-level theory}
\hspace{\parindent}
Beyond tree-level approximation the distribution function for the
inflaton
field $\varphi$ is just the diagonal element of the density matrix
${\rm
Tr}_f|\Psi\!\!><\!\!\Psi|$ obtained from the full quantum state
$|\Psi\!\!>=\Psi(\varphi,f\,|\,t)$ by tracing out the rest of the
degrees of
freedom $f$
	\begin{eqnarray}
	{\mbox{\boldmath $\rho$}}\,(\varphi\,|\,t)=
	\int df\;\Psi^*(\varphi,f\,|\,t)\,\Psi(\varphi,f\,|\,t),
\label{3.1}
	\end{eqnarray}
which does not reduce to a simple squaring of the wavefunction. For
the inner
product in (\ref{3.1}) to be unambiguously defined, the wavefunction
$\Psi(\varphi,f\,|\,t)$ should be taken in the representation of
physical (ADM)
variables with the time $t$ fixed by a chosen ADM reduction procedure
\cite{ADM} (for a perturbative equivalence of the ADM and
Dirac-Wheeler-DeWitt
quantization of gravity see \cite{BPon,BKr,BarvU}). In the
approximation of the
Robertson-Walker model,  the ADM physical variables describing a
spatially
homogeneous background and inhomogeneous field modes (treated
perturbatively) are respectively the inflaton field $\varphi$
and linearized transverse (and traceless) modes $f$ of all possible
spins
\cite{BKam:norm,BarvU,tunnel}, while $t$ can be chosen to be a cosmic
time with
the unit lapse. At late times the interpretation of (\ref{3.1}) can
be obscured
by
subtle issues of decoherence, quantum noise and stochasitcity rapidly
growing
during the inflationary epoch, but at the moment $t=0$ of quantum
nucleation
from the Euclidean spacetime ${\mbox{\boldmath
$\rho$}}\,(\varphi\,|\,0)\equiv{\mbox{\boldmath $\rho$}}\,(\varphi)$
can be
regarded as a distribution function of inflationary universes for
those
$\varphi$ that belong to the domain of the slow-roll approximation .

This quantity was calculated in \cite{BKam:norm,BarvU,tunnel}, and in
this
approximation is given by \cite{BKM}
	\begin{eqnarray}
	{\mbox{\boldmath $\rho$}}_{N\!B,\,T}(\varphi)\cong
	 \frac1{{\mbox{\boldmath $H$}}^2 (\varphi)}\;{\large{\rm e}}^
	{\textstyle\, \mp {\mbox{\boldmath $I$}}(\varphi)-
	{\mbox{\boldmath $\Gamma$}}_{\rm loop}(\varphi)},
\label{3.2}
	\end{eqnarray}
where ${\mbox{\boldmath $I$}}(\varphi)=I(\phi(\varphi))$ is the
Euclidean
action rewritten in the frame of the original Lagrangian
(\ref{2.3}), ${\mbox{\boldmath $H$}}(\varphi)$ is a Hubble constant
in the same
frame related by the equation
${\mbox{\boldmath
$H$}}(\varphi)=H(\phi)\sqrt{1+8\pi|\xi|\,\varphi^2/m_P^2}$ to
the Hubble constant $H(\phi)$ in the Einstein frame (we denote the
quantities
in the original field frame by boldface letters to distinguish them
from those
of the Einstein one).  The main nontrivial ingredient of this
algorithm is the
Euclidean effective action
${\mbox {\boldmath$ \Gamma$}}_{\rm loop}(\varphi)$ accumulating the
quantum
effects of all physical fields $g(x)=(\varphi(x),f(x))$ and beginning
with the
one-loop contribution
	${\mbox {\boldmath$ \Gamma$}}_{\rm 1-loop}(\varphi)=
	\left.(1/2)\,{\rm Tr\,ln}\,
	\delta^2 \!{\mbox{\boldmath $I$}}[\,g\,]/\delta g(x)\delta
g(y)
	\,\right |_{\,\rm \bf D\!S}$.
It is calculated on the quasi-DeSitter gravitational instanton {\bf
DS} -- the
4-dimensional quasi-sphere of radius $1/{\mbox{\boldmath
$H$}}(\varphi)$ --
and,
therefore, parametrized by $\varphi$.

In the high-energy limit of the large inflaton field, including the
slow-roll
domain
and corresponding in the model (\ref{2.3}) to the
Hubble constant ${\mbox{\boldmath $H$}}(\varphi)\simeq
\sqrt{\lambda/12|\xi|}\varphi\rightarrow\infty$
($H(\phi)\rightarrow\sqrt{\lambda/96\pi \xi^2}m_P$  in the Einstein
frame
\cite{SalopBB}), the effective action is calculated on the DeSitter
instanton
of vanishing size ${\mbox{\boldmath $H$}}^{-1}$and, therefore, is
determined
mainly by the total anomalous scaling $Z$ of the theory on such a
manifold
	\begin{eqnarray}
	{\mbox {\boldmath$ \Gamma$}}_{\rm loop}\,\Big|
	_{\,\textstyle{\mbox{\boldmath $H$}}\!\rightarrow\!\infty}
	\simeq Z\,{\rm ln}\,\Big({\mbox{\boldmath $H$}}/\mu\Big),
\label{3.4}
	\end{eqnarray}
where $\mu$ is a renormalization mass parameter or a dimensional
cutoff
generated by the fundamental and finite string theory, if the model
(\ref{2.3})
is regarded as its sub-Planckian effective limit. In the one-loop
approximation
the parameter $Z$ is determined by the total second DeWitt
coefficient
\cite{DW:Dynamical} of all quantum fields  $g=(\varphi,f)$,
integrated over the DeSitter instanton,
	$Z=(1/16\pi^2)\,\int_{\rm \bf D\!S} d^4x\,g^{1/2}a_2(x,x)$
and, thus, crucially depends on the particle content of a model
including as
a graviton-inflaton sector the Lagrangian (\ref{2.3}).

The use of eqs.(\ref{3.2}) and (\ref{3.4}) shows that the quantum
probability
distribution acquires in contrast to its tree-level approximation
(\ref{2.1})
extra $Z$-dependent factor
	\begin{eqnarray}
	{\mbox{\boldmath $\rho$}}_{N\!B,\,T}(\varphi)\cong
	{\rm  exp} \left[\pm
\frac{3m_P^4}{8U(\phi\,(\varphi))}\,\right]
	\,\varphi^{\textstyle -Z\!-\!2},
                \label{3.6}
	\end{eqnarray}
which can make the both no-boundary and tunnelling wavefunctions
normalizable
at over-Planckian scales provided the parameter $Z$ satisfies the
inequality
$Z>-1$ serving as a selection criterion for consistent particle
physics models
with a justifiable semiclassical loop expansion
\cite{BKam:norm,Kam:super}.
Although this equation is strictly valid only in the limit
$\varphi\rightarrow\infty$, it can be used for a qualitatively good
description
at intermidiate energy scales. In this domain the distribution
(\ref{3.6}) can
generate the inflation probability peak at $\varphi=\varphi_I$ with
the
dispersion
$\sigma$, $\sigma^{-2}=-d^2 {\rm ln}\,{\mbox{\boldmath
$\rho$}}(\varphi_I)/
d\varphi_I^2$,
	\begin{eqnarray}
	\varphi_I^2=\frac{2|I_1|}{Z+2},\,\,\,\,\sigma^2=\frac{|I_1|}{(
Z+2)^2},\,\,\,\,
	I_1=-24\pi\frac{|\xi|}{\lambda}\,(1+\delta)\,m_P^2,
       \label{3.7}
	\end{eqnarray}
where $I_1$ is a second coefficient of expansion of the Euclidean
action in
inverse powers of $\varphi$, $ {\mbox{\boldmath $I$}}(\varphi)=
-3m_P^4/8U(\phi\,(\varphi))=I_0+I_1/\varphi^2+O(1/\varphi^4)$. For
the
no-boundary and tunnelling states this peak exists in complimentary
ranges of the parameter $\delta$. For the no-boundary state it can be
realized
only for $\delta<-1\,(I_1>0)$ and, thus, corresponds to the eternal
inflation with the field $\varphi$ on the negative slope of the
inflaton potential (\ref{2.4}) growing from its starting value
$\varphi_I>\bar\varphi$.
For a tunnelling proposal this peak takes place for $\delta\!>\!-1$
and
generates
the finite duration of the inflationary stage with the number of
e-foldings in
the original frame
	${\mbox{\boldmath $N$}}(\varphi_I)=
	(\varphi_I/m_P)^2
\pi(|\xi|+1/6)/(1+\delta)=8\pi^2|\xi|(1+6|\xi|)/\lambda(Z+2)$.
In what follows we consider the latter case because it describes the
conventional scenario with the matter-dominated stage following the
inflation.

\section{Nonminimal inflation and particle physics of the early
Universe}
\hspace{\parindent}
The physics of the inflationary Universe with nonminimal inflaton
field
(\ref{2.3}) was recently analyzed in much detail in \cite{SalopBB}
where it was
persuasively advocated that this model generates the spectrum of
density
perturbations compatible with COBE measurements \cite{Salopek} in the
range of
coupling constants $\lambda/\xi^2\sim 10^{-10}$ (the experimental
bound on the
gauge-invariant \cite{BST} density perturbation $P_{\zeta}(k)=N^2_k
(\lambda/\xi^2)/8\pi^2$ in the $k$-th mode "crossing" the horizon at
the moment
of the e-foldings number $N_k$).  This allows one to avoid the
unnaturally
small value of $\lambda$ in the minimal inflaton model \cite{Linde}
and replace
it with the GUT compatible value $\lambda\simeq 0.05$, provided
$\xi\!\simeq\!
-2\!\times\! 10^4$ is chosen to be related to the ratio of the Planck
scale to
a
typical GUT scale, $|\xi|\sim m_P/v$. For these coupling constants
the bound
${\mbox{\boldmath $N$}}(\varphi_I)\geq 60$ on the duration of the
inflation, generated by the probability peak (\ref{3.7}), results in
an
enormous value of the anomalous scaling $Z\sim 10^{11}$. A remarkable
feature
of the proposed scheme is that this huge value can be naturally
induced by
large $\xi$ already in the one-loop approximation. Indeed, the
expression for
$Z_{1-\rm loop}$, well known for a generic theory
\cite{DW:Dynamical}, has a
contribution quartic in effective masses of  physical particles
easily
calculable on a spherical DeSitter background \cite{Al-FrTs}
 $Z_{1-\rm loop}=(12{\mbox{\boldmath
$H$}}^4)^{-1}(\sum_{\chi}m_{\chi}^4
+4\sum_{A}m_{A}^4-4\sum_{\psi}m_{\psi}^4)+...$,
where the summation goes over all Higgs scalars $\chi$, vector gauge
bosons $A$
and Dirac spinors $\psi$. Their effective masses for large $\varphi$
are
dominated by the contributions
$m_{\chi}^2=\lambda_{\chi}\varphi^2/2,\,m_{A}^2=g_{A}^2\varphi^2$ and
$m_{\psi}^2=f_{\psi}^2\varphi^2$ induced via the Higgs mechanism from
their
interaction Lagrangian with the inflaton field
	\begin{eqnarray}
	{\mbox{\boldmath $L$}}_{\rm
int}=\sum_{\chi}\frac{\lambda_{\chi}}4
	\chi^2\varphi^2
	+\sum_{A}\frac12 g_{A}^2A_{\mu}^2\varphi^2+
	\sum_{\psi}f_{\psi}\varphi\bar\psi\psi + {\rm derivative\,\,
coupling}.
\label{4.1}
	\end{eqnarray}
Thus, in view of the relation $\varphi^2/{\mbox{\boldmath
$H$}}^2=12|\xi|/\lambda$, we get the leading contribution of large
$|\xi|$ to
the total anomalous scaling of the theory
	\begin{eqnarray}
	Z_{\rm 1-loop}=6\,\frac{\xi^2}{\lambda}{\mbox{\boldmath
$A$}}+O(\xi), \,\,\,
	{\mbox{\boldmath
$A$}}=\frac1{2\lambda}\Big(\sum_{\chi}\lambda_{\chi}^2
	+16\sum_{A}g_{A}^4-16\sum_{\psi}f_{\psi}^4\Big) ,
\label{4.2}
	\end{eqnarray}
which contains the same large dimensionless ratio
$\xi^2/\lambda\simeq 10^{10}$
and the universal quantity ${\mbox{\boldmath $A$}}$ determined by a
particle
physics model (gravitons and inflaton field do not contribute to
${\mbox{\boldmath $A$}}$, as well as gravitino in case when the
latter is
decoupled from the inflaton).

For such $Z_{\rm 1-loop}$ the parameters of the inflationary peak
express as
	\begin{eqnarray}
	\varphi_I=m_P\sqrt{\frac{8\pi(1+\delta)}{|\xi|\,
	{\mbox{\boldmath $A$}}}},\,\,\,\,
	\sigma=\frac{\varphi_I}{\sqrt{12{\mbox{\boldmath $A$}}}}
	\frac{\sqrt{\lambda}}{|\xi|}, \,\,\,\,
	{\mbox{\boldmath $H$}}(\varphi_I)=\sqrt{|\xi|}\sigma,
\,\,\,\,
	{\mbox{\boldmath $N$}}(\varphi_I)=
	\frac{8\pi^2}{\mbox{\boldmath $A$}}
\label{4.3}
	\end{eqnarray}
and satisfy the bound  ${\mbox{\boldmath $N$}}(\varphi_I)\geq 60$
with a single
restriction on ${\mbox{\boldmath $A$}}$, ${\mbox{\boldmath
$A$}}\leq1.3$.
This restriction
simultaneously provides a very good slow-roll approximation, because
the
corresponding smallness parameter (in the original frame of the
Lagrangian
(\ref{2.3})) is
$\dot\varphi/{\mbox{\boldmath $H$}}\varphi\simeq-
{\mbox{\boldmath $A$}}/96\pi^2
\sim-10^{-3}$. For a value of $\delta \ll1$ ($\delta\sim 8\pi/|\xi|$
for
$|\xi|\sim m_P/v$) and ${\mbox{\boldmath $A$}}\simeq 1$,  the
obtained
numerical parameters describe extremely sharp inflationary peak at
$\varphi_I\simeq 0.03 m_P$ with small width $\sigma\simeq 10^{-7}m_P$
and
sub-Planckian Hubble constant ${\mbox{\boldmath
$H$}}(\varphi_I)\simeq
10^{-5}m_P$, which is exactly the most realistic range of
inflationary
scenario.  The smallness of the width does not, however, lead to its
quick
quantum spreading: the commutator relations for operators
${\hat\varphi}$ and
$\dot{\hat\varphi}$, $[{\hat\varphi},\dot{\hat\varphi}]\simeq
i/(12\pi^2|\xi|a^3)$ \cite{SalopBB}, give rise at the beginning of
the
inflation, $a\simeq {\mbox{\boldmath $H$}}^{-1}$, to a negligible
dispersion of
$\dot\varphi$, $\Delta\dot\varphi\simeq {\mbox{\boldmath
$H$}}^3/12\pi^2|\xi|\sigma
\simeq(8/{\mbox{\boldmath
$A$}})(\sqrt{\lambda}/|\xi|)|\dot\varphi|\ll|\dot\varphi|$. It is
remarkable
that the relative width
$\sigma/\varphi_I\sim\Delta{\mbox{\boldmath $H$}}
/{\mbox{\boldmath $H$}}\sim 10^{-5}$ corresponds to the observable
level of
density perturbations, although it is not clear whether this quantum
dispersion
$\sigma$ is directly measurable now, because of the stochastic noise
of the
same order of magnitude generated during the inflation and
superimposed upon
$\sigma$.

All these conclusions are rather universal and (apart from the choice
of
$|\xi|$ and $\lambda$) universally depend on one parameter
${\mbox{\boldmath
$A$}}$ (\ref{4.2}) of the particle physics model.  This quantity
should satisfy
the bound
	\begin{eqnarray}
	0<{\mbox{\boldmath $A$}}\leq 1.3              \label{4.4}
	\end{eqnarray}
in order to render $Z$ positive, thus suppressing over-Planckian
energy
scales, and  provide sufficient amount of inflation
(${\mbox{\boldmath $A$}}$
should not, certainly, be exeedingly close to zero, not to suppress
the
dominant
contribution of large $|\xi|$ in (\ref{4.2})). This bound again
suggests the
quasi-supersymmetric nature of the particle model, although for
reasons
different from the conclusions of
\cite{Kam:super}.  It is only supersymmetry that can constrain the
values of
the Higgs $\lambda_{\chi}$, vector gauge $g_A$ and Yukawa $f_{\psi}$
couplings
so as to provide a subtle balance between the contributions of bosons
and
fermions in (\ref{4.2}) and fit the quantity
${\mbox{\boldmath $A$}}$ into a narrow range (\ref{4.4}).  In
contrast to
ref.\cite{Kam:super}, this conclusion is robust against the
subtleties of the
definition of $Z$ (related to the treatment of zero modes on DeSitter
background \cite{Al-FrTs}) because it probes only the large limit of
$Z\gg 1$.

\section{Conclusions}
\hspace{\parindent}
Thus, the same mechanism that suppresses the over-Planckian
energy scales also generates a narrow probability
peak in the distribution of tunnelling inflationary universes and
strongly
suggests the (quasi)supersymmetric nature of their particle content.
It
seems to be remarkably consistent with microwave background
observations
within the model with a strongly coupled nonminimal inflaton field.
This result
is
independent of the renormalization ambiguity, which gives a hope that
it is
also robust against  inclusion of multi-loop corrections. It is usual
to be
prejudiced against a large value of the nonminimal coupling $|\xi|$
which
generates
large quantum effects leaving them uncontrollable in multi-loop
orders.  This
is not, however, quite correct, because the effective gravitational
constant in
such a model is inverse proportional to $m_P^2+8\pi|\xi|\varphi^2$
and, thus,
large $|\xi|$ might improve the loop expansion \cite{BKK}. Obviously,
the large
value of $|\xi|$ at sub-Planckian (GUT) scale requires explanation
which might
be based on the renormalization group approach (and its extension to
non-renormalizable theories \cite{BKK}). As shown in \cite{BKK},
quantum
gravity with nonminimal scalar field has an asymptotically free
conformally
invariant ($\xi=1/6$) phase at over-Planckian regime, which is
unstable at
lower energies. It is plausible to conjecture that this instability
can lead
(via composite states of the scalar field) to the inversion of the
sign of
running $\xi$ and its growth at the GUT scale, thus making possible
the
proposed inflation applications \cite{BK}.

As far as it concerns the GUT and lower energy scales,  the ground
for
supersymmetry of the above type looks
very promising in the context of a special property of supersymmetric
models to
have a single unification point for weak, electromagnetic and strong
interactions
(the fact that has been discovered in 1987 and now becoming widely
reckognized
after the recent experiments at LAP \cite{supersym}). The proposed
quantum
gravitational implication of supersymmetry might also be useful in
view of the
raising interest in the supersymmetric analogues of the
Wheeler-DeWitt
equations
and their non-perturbative solutions \cite{D'Eath et al}.

{}From the viewpoint of the theory of the early universe and its
observational
status, the obtained results give a strong preference to the
tunnelling
cosmological wavefunction. This argument, however, can hardly be
conclusive, because the difference between the no-boundary and
tunnelling wavefunctions might be ascribed to the open problem of the
correct quantization of the conformal mode. Note that the
normalizability
criterion for the distribution function  and its algorithm
(\ref{3.2}) do not
extend to the low-energy limit $\varphi\rightarrow 0$, where the
naively
computed no-boundary distribution function blows up to infinity, the
slow-roll
approximation becomes invalid, etc. This is a domain related to a
highly
speculative (but, probably, inevitable) third quantization of gravity
\cite{Coleman}, which goes beyond the scope of this paper.
Fortunately, this
domain is separated from the obtained inflationary peak by a vast
desert with
practically zero density of the quantum distribution, which
apparently
justifies our conclusions disregarding the ultra-infrared physics of
the
Coleman theory of baby universes and cosmological constant
\cite{Coleman}.

\section*{Acknowledgements}
\hspace{\parindent}
The authors benefitted from helpful discussions with Don N.Page and
D.Salopek.
The work of A.O.B. on this paper was supported by NSERC grant at the
University of Alberta.  A.Yu.K. is grateful to International Science
Foundation
for
financial support and to Russian Foundation for Fundamental
Researches
for support under grant No 94-02-03850-a.

\end{document}